# Systematic KMTNet Planetary Anomaly Search. IV. Complete Sample of 2019 Prime-Field

Weicheng Zang[1,A ★] , Hongjing Yang[1,A] , Cheongho Han[2,A], Chung-Uk Lee[3,A], Andrzej Udalski[4,B],
Andrew Gould[5,6,A], Shude Mao[1,7] , Xiangyu Zhang[5], Wei Zhu[1], Michael D. Albrow[8,A], Sun-Ju Chung[3,9,A],
Kyu-Ha Hwang[3,A], Youn Kil Jung[3,A], Yoon-Hyun Ryu[3,A], In-Gu Shin[3,A], Yossi Shvartzvald[10,A],
Jennifer C. Yee[11,A], Sang-Mok Cha[3,12,A], Dong-Jin Kim[3,A], Hyoun-Woo Kim[3,A], Seung-Lee Kim[3,9,A],
Dong-Joo Lee[3,A], Yongseok Lee[3,12,A], Byeong-Gon Park[3,9,A], Richard W. Pogge[6,A], Przemek Mróz[4,B] ,
Jan Skowron[4,B] , Radoslaw Poleski[4,B] , Michał K. Szymański[4,B] , Igor Soszyński[4,B] ,
Paweł Pietrukowicz[4,B] , Szymon Kozłowski[4,B] , Krzysztof Ulaczyk[13,B] , Krzysztof A. Rybicki[4,B] ,
Patryk Iwanek[4,B] , Marcin Wrona[4,B] , Mariusz Gromadzki[4,B]

[1] Department of Astronomy, Tsinghua University, Beijing 100084, China
[2] Department of Physics, Chungbuk National University, Cheongju 28644, Republic of Korea
[3] Korea Astronomy and Space Science Institute, Daejon 34055, Republic of Korea
[4] Astronomical Observatory, University of Warsaw, Al. Ujazdowskie 4, 00-478 Warszawa, Poland
[5] Max-Planck-Institute for Astronomy, Königstuhl 17, 69117 Heidelberg, Germany
[6] Department of Astronomy, Ohio State University, 140 W. 18th Ave., Columbus, OH 43210, USA
[7] National Astronomical Observatories, Chinese Academy of Sciences, Beijing 100101, China
[8] University of Canterbury, Department of Physics and Astronomy, Private Bag 4800, Christchurch 8020, New Zealand
[9] University of Science and Technology, Korea, (UST), 217 Gajeong-ro Yuseong-gu, Daejeon 34113, Republic of Korea
[10] Department of Particle Physics and Astrophysics, Weizmann Institute of Science, Rehovot 76100, Israel
[11] Center for Astrophysics | Harvard & Smithsonian, 60 Garden St.,Cambridge, MA 02138, USA
[12] School of Space Research, Kyung Hee University, Yongin, Kyeonggi 17104, Republic of Korea
[13] Department of Physics, University of Warwick, Gibbet Hill Road, Coventry CV4 7AL, UK
[A] The KMTNet Collaboration
[B] The OGLE Collaboration

20 July 2022

**ABSTRACT**

We report the complete statistical planetary sample from the prime fields ($\Gamma \geqslant 2$ hr$^{-1}$) of the 2019 Korea Microlensing Telescope Network (KMTNet) microlensing survey. We develop the optimized KMTNet AnomalyFinder algorithm and apply it to the 2019 KMTNet prime fields. We find a total of 13 homogeneously selected planets and report the analysis of three planetary events, KMT-2019-BLG-(1042,1552,2974). The planet-host mass ratios, $q$, for the three planetary events are $6.34 \times 10^{-4}$, $4.89 \times 10^{-3}$ and $6.18 \times 10^{-4}$, respectively. A Bayesian analysis indicates the three planets are all cold giant planets beyond the snow line of their host stars. The 13 planets are basically uniform in $\log q$ over the range $-5.0 < \log q < -1.5$. This result suggests that the planets below $q_{\rm break} = 1.7 \times 10^{-4}$ proposed by the MOA-II survey may be more common than previously believed. This work is an early component of a large project to determine the KMTNet mass-ratio function, and the whole sample of 2016–2019 KMTNet events should contain about 120 planets.

## 1 INTRODUCTION

Microlensing is currently the only method that can detect cold planets with masses down to Earth mass (Mao & Paczynski 1991; Gould & Loeb 1992; Mao 2012; Gaudi 2012), but the number of detections ($\sim 130$) has increased slowly compared to those of the transit ($\sim$

3900) and the radial velocity ($\sim 900$) methods[1]. There are two main challenges in microlensing planetary searches. First, the optical depth to microlensing is only $\tau \sim 10^{-6}$ (Sumi et al. 2013; Mróz et al. 2020), and the probability of detecting a planet within a given microlensing event is roughly $q^{1/2}$, where $q$ is the planet-to-host mass ratio. Thus, a large-area survey toward a region with high stellar number density [1] http://exoplanetarchive.ipac.caltech.edu as of 2022 April 1st.

(e.g., the Galactic bulge) is needed to detect enough microlensing events for planetary searches. Second, because the typical duration of microlensing planetary signal is short ($\lesssim 1$ day), high-cadence observations are required ($\Gamma \sim 1 \, \mathrm{hr}^{-1}$ for "Neptunes" and $\Gamma \sim 4 \, \mathrm{hr}^{-1}$ for "Earths", Henderson et al. 2014).

The ideal mode (for ground-based observations) is to conduct continuous, wide-area, high-cadence surveys from multiple sites. However, due to the scarcity of telescope resources, microlensing planets before 2016 were discovered mainly using the strategy advocated by Gould & Loeb (1992) (see Figure 10 of Mróz et al. 2017), which is a combination of wide-area, low-cadence surveys for finding microlensing events and intensive follow-up observations for capturing the planetary perturbation (e.g., Udalski et al. 2005). This mode yielded two homogeneous samples, but their sizes are small. Gould et al. (2010) found six planets from a well-defined sample of 13 high-magnification events intensively observed by the Microlensing Follow Up Network ($\mu$FUN) during 2005–2008. Cassan et al. (2012) contains three planets from the 196 events observed by the PLANET follow-up network during 2002–2007.

The second phase of Microlensing Observations in Astrophysics (MOA-II, 2006+) was the first to conduct wide-area, high-cadence surveys toward the Galactic bulge, using its one 1.8 m telescope in New Zealand and equipped with 2.2 deg$^2$ camera (Sako et al. 2008). The fourth phase of the Optical Gravitational Lensing Experiment (OGLE-IV) joined the wide-area, high-cadence surveys in 2011 using its one 1.3 m telescope equipped with 1.4 deg$^2$ camera in Chile (Udalski et al. 2015). The Wise microlensing survey, with its one 1.0 m telescope equipped with its 1.0 deg$^2$ camera in Israel, joined the global network during 2011–2014 (Shvartzvald et al. 2016). By combining the OGLE, MOA, and Wise surveys, they were able to conduct a continuous, high-cadence microlensing survey for 8 deg$^2$ of the Galactic bulge. From their observations, there are three homogeneous samples. Suzuki et al. (2016) analyzed 1474 MOA-II events and found 23 planets ($q < 0.03$). Shvartzvald et al. (2016) detected nine planets ($\log q < -1.4$) from 224 events observed by OGLE-IV, MOA-II, and Wise. Poleski et al. (2021) studied the six wide-orbit ($s > 2$, where $s$ is the planet-host separation in units of the Einstein radius $\theta_E$) planets in a sample of 3112 OGLE events. Two of these samples are larger than the two Gould & Loeb (1992) strategy samples, but they contain relatively few low mass-ratio planets ($q < 10^{-4}$) planets, with just two in the Suzuki et al. (2016) sample and one in the Shvartzvald et al. (2016) sample.

The ideal mode of microlensing planetary searches was realized by the new-generation microlensing survey, the Korea Microlensing Telescope Network (KMTNet, Kim et al. 2016), which consists of three 1.6 m telescopes equipped with 4 deg$^2$ cameras in Chile (KMTC), South Africa (KMTS), and Australia (KMTA). Beginning in 2016, KMTNet conducted near-continuous observations for a total of (3, 7, 11, 3) fields at cadences of $\Gamma_K \sim (4, 1, 0.4, 0.2) \, \mathrm{hr}^{-1}$ (see Figure 12 of Kim et al. 2018b). The advent of KMTNet greatly increased the microlensing planet detection number, with about half of all published microlensing planets having been detected with the KMTNet data (see Figure 1 of Zang et al. 2021). However, almost all of the KMTNet planets published before 2021 were found using by-eye searches, which demonstrably missed some subtle planetary signals and thus do not constitute a large-scale homogeneous sample for statistical studies.

To exhume the buried planetary signals in KMTNet data, Zang et al. (2021) developed the KMTNet AnomalyFinder algorithm. This algorithm revealed many planetary signals that were missed in previous by-eye searches. Zang et al. (2021) and Hwang et al. (2022) found a total of 11 planets with $q < 2 \times 10^{-4}$, including seven whose planetary signals had not been noticed before. Wang et al. (2022) discovered the lowest-$q$ microlensing planet at $s > 2$.

More importantly, this approach opens the door to a large-scale homogeneous KMTNet sample for statistical studies. Our ultimate goal is to measure the KMTNet mass-ratio function for the full planetary range ($q < 0.03$) with at least 2016–2019 KMTNet data. To achieve this goal, we are undertaking work in five areas. First, we need an optimized KMTNet AnomalyFinder algorithm, because the main purpose of the work by Zang et al. (2021) was to develop and test the method and programming. Second, we must apply it to at least the 2016–2019 KMTNet data. Third, a KMTNet quasi-automated tender-loving care (TLC) re-reductions pipeline is needed to make the large-scale intensive re-reductions tractable. H. Yang et al. (in prep) has built this pipeline by adapting the pySIS package (Albrow et al. 2009), by which each event only needs about one-hour of human effort on average. Fourth, a planet detection efficiency calculator based on the optimized KMTNet AnomalyFinder algorithm is needed. Jung et al. (in prep) has realized the calculator. Fifth, all candidate planetary events identified by the optimized KMTNet AnomalyFinder algorithm should be fitted to determine their mass ratios. This aspect requires intensive effort because KMTNet data annually yield about 80 candidate planetary events. We thus group and analyze annual candidate planetary events according to their KMTNet observing cadences, with $\Gamma_K \geqslant 2 \, \mathrm{hr}^{-1}$ for the prime-field sample and $\Gamma_K < 2 \, \mathrm{hr}^{-1}$ for the sub-prime-field sample.

In the present paper, we introduce the 2019 KMTNet prime-field planetary sample. We first describe the optimized KMTNet Anomaly-Finder algorithm in Section 2. We then introduce AnomalyFinder results for the 2019-Prime Fields and the protocols for our planetary sample in Section 3. We present the observations, the light-curve analysis, and the physical parameters for three unpublished planets in the 2019 prime-field sample in Sections 4, 5 and 6, respectively. Finally, we discuss the implications from the 2019 KMTNet prime-field planetary sample in Section 7.

## 2 THE OPTIMIZED KMTNET ANOMALYFINDER ALGORITHM

The detailed descriptions of the KMTNet AnomalyFinder algorithm were presented in Section 2 of Zang et al. (2021). Its basic idea is to search for anomalies from the residuals to a best-fit point-source point-lens (PSPL, Paczyński 1986) model, and its basic algorithm is based on the KMTNet EventFinder algorithm (Kim et al. 2018b), which adopts a Gould (1996) 2-dimensional grid of ($t_0, t_{\mathrm{eff}}$) to fit anomalies, where $t_0$ is the time of maximum magnification, and $t_{\mathrm{eff}}$ is the effective timescale. The ultimate KMTNet AnomalyFinder algorithm used for the KMTNet statistical sample contains two main improvements.

### 2.1 Data Handling

Because "bad" points frequently produce fake anomalies, Zang et al. (2021) aggressively removed all data points that have a sky background brighter than 5000 ADU/pixel[2] or a seeing FWHM larger than 7 pixels (0.4″ per pixel) for the KMTA and KMTS data and 6.5 pixels for the KMTC data. However, many data points above these thresholds are of good quality and could contribute to the

---

identification of anomalies (e.g., the planetary signal of KMT-2018-BLG-0029, Gould et al. 2020). We also find that there is no clear relation between fake anomalies and sky background. Thus, we group all the data points as "good" seeing (FWHM < 7 pixels for KMTA and KMTS, and FWHM < 6.5 pixels for KMTC), "ok" seeing (7 ⩽ FWHM < 9 pixels for KMTA and KMTS, and 6.5 ⩽ FWHM < 9 pixels for KMTC), and "bad" seeing (FWHM > 9 pixels). For the PSPL fits, we only use the "good" points, but all points are shown to the operator.

In general, the errors derived from photometric pipelines, are underestimated and must be renormalized using the method of Yee et al. (2012), which enables, for each group of data, $\chi^2$/dof to become unity and the cumulative sum of $\chi^2$ are approximately linear as a function of source magnification, where dof is the degree of freedom. However, because the online data used in the search contain many outliers (mainly "ok" and "bad" points), the Yee et al. (2012) method is infeasible. For each event, Zang et al. (2021) simply multiplied each error by $k$, and $k = 1.5$. For the optimized algorithm, we first follow Kim et al. (2018b) in removing the 10% worst-$\chi^2$ points and then calculate $k$ for each group of data by keeping $\chi^2$/dof = 1 for the remaining data.

### 2.2 Manual Review

Each candidate identified by the automated algorithm is shown to the operator (W. Zang), and the operator then selects the plausible signals. See Figure 1 for an example. Zang et al. (2021) provided a four-panel display to the operator, including the light curves and residuals for the candidate signal and the data of the whole season. The optimized algorithm adds a panel that shows the seeing information for the candidate signals. This new panel shows the operator whether a candidate signal has a clear trend with seeing and whether "ok" and "bad" points dominate the detection. In addition, the optimized algorithm bins the residual of the whole season by 0.5 days, which enables the operator to see more clearly whether the data are variable and whether there is a buried long-duration trend for a short-timescale event (e.g., KMT-2018-BLG-0998, Wang et al. 2022).

Zang et al. (2021) only tentatively classified the plausible signals by eye as "possible planetary events" and "other anomalies". The optimized algorithm allows them to be classified by eye into one of five categories, "planet", "planet/binary" (more likely a planet), "binary/planet" (more likely a binary), "binary", and "finite source" (finite-source point-lens Gould 1994; Witt & Mao 1994; Nemiroff & Wickramasinghe 1994). In principle, each of plausible signals should be fitted to determine its $q$, but the annual KMTNet data produces about 2000 plausible signals, which is a heavy burden. Thus, the operator has to identify unambiguously non-planetary events (i.e., "binary") by eye, and we only fit the other four categories. There is a risk that the operator misidentifies a planetary event as "binary". However, co-author C. Han has personal modeling for ∼ 75% of the "binary". We cross-checked his modeling for 2016–2019 "binary" and found that the operator only missed one massive ($q \gtrsim 0.001$) planet on average per year. Thus, the potential buried planets in the remaining about ∼ 25% of the "binary" can be ignored compared to the roughly 30–50 planets from the annual KMTNet data. Even so, systematic modeling and analysis of all "binary" events are needed in the future. This work will not only complete the KMTNet planetary sample from AnomalyFinder, but also exhume buried planets in binary systems. Indeed, a recent study (Kuang et al. 2022) found that the current KMTNet sample for planets in binary systems is incomplete, which is likely due to the deficiency of the current KMTNet

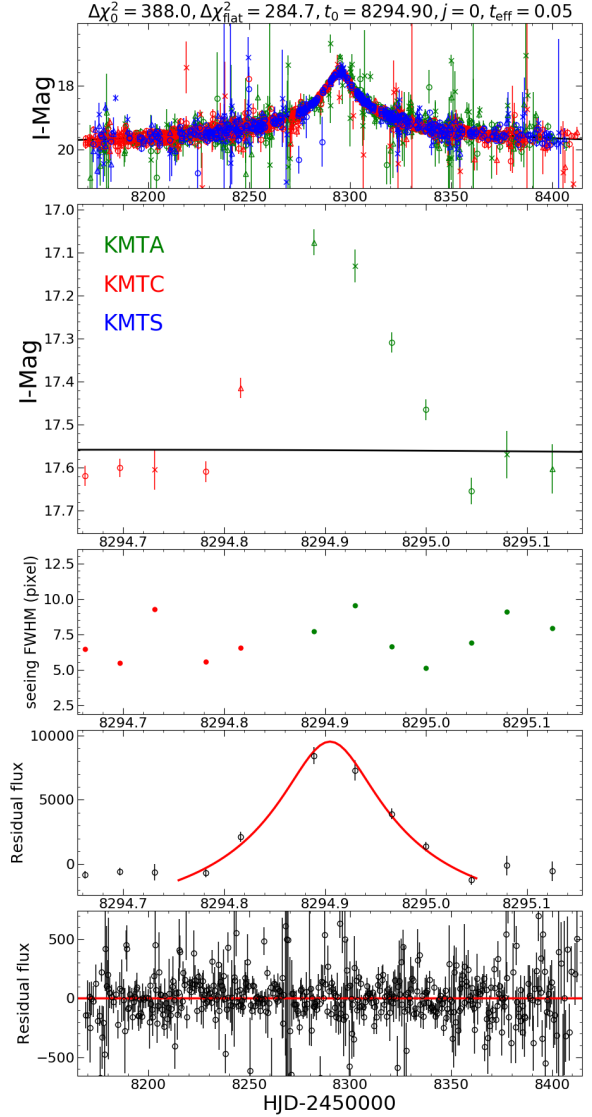

**Figure 1.** Example of the planetary signal of KMT-2018-BLG-0029 (Gould et al. 2020) as shown to the operator. Compared with the four-panel display from the original AnomalyFinder (see Figure 2 of Zang et al. 2021), the optimized algorithm has three improvements. First, it adds a panel (the third panel) to show the seeing information for the candidate signals. Second, it bins the residual of the whole season by 0.5 days, shown on the bottom panel. Third, it includes "ok" and "bad" points and plots them using the up triangle and "×" markers, respectively. Without the "ok" and "bad" points, there is only one KMTA point at HJD′ = 8294.96 significantly deviated from the PSPL model, so the operator of the original AnomalyFinder did not select this signal. Although "ok" and "bad" points dominate this detection, this signal does not have a clear trend with seeing, and the KMTA and the KMTC data together form a short-lived bump, so the operator of the optimized AnomalyFinder classified it into the "planet" category (Jung et al. in prep).

AnomalyFinder and by-eye searches. Before systematic modeling and analysis of all "binary" events are completed, the same operator (W.Zang) would have to perform the manual review to ensure the current misclassification rate in the future. In addition, for the other four categories for which we fit all events, we track the operator's by-eye sensitivity about planetary events.

## 3 ANOMALYFINDER RESULTS FOR THE 2019-PRIME FIELDS AND PROTOCOLS

For the optimized algorithm, the lower limit of $t_{\rm eff}$ is reduced to 0.05 days. For the prime-field sample, the criteria are $\Delta\chi_0^2 > 200$, or $\Delta\chi_0^2 > 120$ and $\Delta\chi_{\rm flat}^2 > 60$. See Zang et al. (2021) for their detailed definitions. As a result, there are 14007 candidate signals from 883 events. The operator identified 9 "planet", 8 "planet/binary", 15 "binary/planet", 83 "binary", and 7 "finite source".

Among them, six planets have been identified using by-eye searches, including four that were previously published (KMT-2019-BLG-0842, Jung et al. 2020b; KMT-2019-BLG-1953, Han et al. 2020; OGLE-2019-BLG-0954, Han et al. 2021b; KMT-2019-BLG-1715, Han et al. 2021a), one that is in preparation (OGLE-2019-BLG-1180, Chung in prep) and one that is analyzed in this paper (KMT-2019-BLG-1552). For the remaining events from the four categories, the operator checked whether the OGLE and/or MOA data points are inconsistent with the KMTNet-based anomalies, cross-checked with C. Han's modeling, and fitted the candidate planetary events to binary-lens single-source (2L1S, Gaudi 1998) model using the online data. At this stage, we did not consider the potential degeneracy from the single-lens binary-source (1L2S, Gaudi 1998) model. There are 14 events that required TLC reductions. Of these, one needs TLC reductions to confirm the anomaly, and 13 are potentially planetary with $q_{\rm online} < 0.05$. We expect that this sample will be complete for at least $q < 0.03$, because we find that the TLC reduction data and online data result in similar $q$ ($< 50\%$ difference) for most events. With the TLC reduction data, four were previously published planetary events with $q < 2 \times 10^{-4}$ (OGLE-2019-BLG-1053, Zang et al. 2021; KMT-2019-BLG-0253, Hwang et al. 2022; OGLE-2019-BLG-1492, Hwang et al. 2022; KMT-2019-BLG-0953, Hwang et al. 2022), three were found to have $0.03 < q < 0.06$ (KMT-2019-BLG-0814/OGLE-2019-BLG-0733 has $q = 0.050$, OGLE-2019-BLG-1067/KMT-2019-BLG-1498 has $q = 0.047$, and KMT-2019-BLG-3301 has $q = 0.053$), four were found to have $q < 0.03$ with unambiguous mass-ratio determinations (KMT-2019-BLG-1042, KMT-2019-BLG-2974, OGLE-2019-BLG-0954, KMT-2019-BLG-1470). In the next three sections, we present the analysis of the three $q < 0.03$ events, KMT-2019-BLG-1042, KMT-2019-BLG-1552 and KMT-2019-BLG-2974, and the analysis of OGLE-2019-BLG-0954 and KMT-2019-BLG-1470 will be presented elsewhere.

In total, there are 14 planets found by the optimized KMTNet AnomalyFinder algorithm in the 2019 KMTNet prime fields. We note that the old AnomalyFinder algorithm identified all of the 14 planetary events, but the operator did not select KMT-2019-BLG-2974 due to the absence of the seeing information, which illustrates the importance of the seeing information in Manual Review.

## 4 OBSERVATIONS AND DATA REDUCTION

The first two events, KMT-2019-BLG-1042 and KMT-2019-BLG-1552, were both first discovered by the KMTNet alert-finder system (Kim et al. 2018a) in 2019. KMT-2019-BLG-1552 was then independently discovered by the OGLE Early Warning System (Udalski et al. 1994; Udalski 2003) about 14 days after KMTNet's alert and designated as OGLE-2019-BLG-1142. The third event, KMT-2019-BLG-2974, was found by the KMTNet post-season EventFinder algorithm (Kim et al. 2018b). The three events all lie in two overlapping KMTNet fields, with a combined cadence of $\Gamma_K \sim 4$ hr$^{-1}$. KMT-2019-BLG-2974 was located in the OGLE BLG504 field, with a cadence of $\Gamma_O \sim 1$ hr$^{-1}$. For both surveys, the great majority of images were taken in the $I$ band, with 9% of KMTNet data and occasional OGLE data taken in the $V$ band for source color measurements.

The data used in the light curve analysis were reduced using the re-reductions pipelines based on the difference image analysis technique (Tomaney & Crotts 1996; Alard & Lupton 1998): TLC pySIS (Albrow et al. 2009) pipeline for the KMTNet data and Wozniak (2000) pipeline for the OGLE data. For the KMTNet data of each event, we conducted pyDIA photometry (Albrow 2017) to measure the source color, which simultaneously yields the light curve on the same magnitude system as field-star photometry. We summarize the basic observational information for the three events in Table 1.

## 5 LIGHT-CURVE ANALYSIS

### 5.1 Preamble

We begin by introducing the processes common to light curve analysis for the three events. Their light curves all show deviations from the normal PSPL model characterized by three parameters: $(t_0, u_0, t_{\rm E})$, i.e., the time of the closest lens-source alignment, the impact parameter scaled to the angular Einstein radius $\theta_{\rm E}$, and the timescale required to cross the unit Einstein radius,

$$t_{\rm E} = \frac{\theta_{\rm E}}{|\vec{\mu}_{\rm rel}|}; \qquad \theta_{\rm E} = \sqrt{\kappa M_{\rm L}\pi_{\rm rel}}; \qquad \kappa \equiv \frac{4G}{c^2{\rm au}} \approx 8.144\frac{\rm mas}{M_\odot}, \quad (1)$$

where $M_{\rm L}$ is the mass of the lens system and $(\pi_{\rm rel}, \vec{\mu}_{\rm rel})$ are the lens-source relative (parallax, proper motion).

We first search for the 2L1S model of each event. The 2L1S model requires three additional parameters $(s, q, \alpha)$ to describe the binary geometry, i.e., the projected separation between the binary components normalized to $\theta_{\rm E}$, the mass ratio between the binary components, and the angle between the source trajectory and the binary axis. We also introduce the parameter $\rho$, the angular source radius $\theta_*$ in units of $\theta_{\rm E}$ ($\rho = \theta_*/\theta_{\rm E}$), for finite-source effects. We use the advanced contour integration code (Bozza 2010; Bozza et al. 2018), VBBinaryLensing[3], to compute the 2L1S magnification $A(t)$ at any given time $t$. In addition, for each data set $i$, there are two linear parameters $(f_{{\rm S},i}, f_{{\rm B},i})$ representing the flux of the source star and any blend flux, respectively. The observed flux, $f_i(t)$, is modeled as

$$f_i(t) = f_{{\rm S},i}A(t) + f_{{\rm B},i}. \qquad (2)$$

To locate the local minima of the 2L1S model, we carry out grid searches for the parameters $(\log s, \log q, \alpha, \rho)$. For each event, we first conduct a sparse grid, which consists of 21 values equally spaced between $-1.0 \leqslant \log s \leqslant 1.0$, 20 values equally spaced between $0° \leqslant \alpha < 360°$, 61 values equally spaced between $-6.0 \leqslant \log q \leqslant 0.0$ and five values equally spaced between $-3.5 \leqslant \log \rho \leqslant -1.5$. For each grid, we find the minimum $\chi^2$ by Markov chain Monte Carlo (MCMC) $\chi^2$ minimization using the emcee ensemble sampler (Foreman-Mackey et al. 2013), with fixed $\log q$, $\log s$ and free $t_0, u_0, t_{\rm E}, \rho, \alpha$. For each local minimum, we then conduct a denser grid search (e.g., Zang et al. 2020). Finally, we find the best-fit models by MCMC with all parameters free.

Gaudi (1998) pointed out that a 1L2S event can mimic a 2L1S event when the second source is much fainter than the primary source and moves much closer to the lens, and there have been several events with plausible 2L1S planetary solutions that proved to be 1L2S events (Hwang et al. 2013; Jung et al. 2017; Rota et al. 2021) or even a single-lens triple-source (1L3S) event (Hwang et al. 2018). Thus, we check whether the observed data can be interpreted by the 1L2S model. The total magnification of a 1L2S model is the superposition of two single-lens events,

$$A_\lambda = \frac{A_1 f_{1,\lambda} + A_2 f_{2,\lambda}}{f_{1,\lambda}} = \frac{A_1 + q_{f,\lambda} A_2}{1 + q_{f,\lambda}}; \quad q_{f,\lambda} \equiv \frac{f_{2,\lambda}}{f_{1,\lambda}}. \quad (3)$$

---

[3] http://www.fisica.unisa.it/GravitationAstrophysics/VBBinaryLensing.htm

**Table 1.** Basic observational information for the three planetary events

| Event Name | KMT-2019-BLG-1042 | KMT-2019-BLG-1552 /OGLE-2019-BLG-1142 | KMT-2019-BLG-2974 |
|---|---|---|---|
| Alert Date | 03 June 2019 | 08 Jul 2019 | Post Season |
| $RA_{J2000}$ | 18:02:10.57 | 17:58:22.99 | 17:51:24.50 |
| $Decl._{J2000}$ | $-27$:34:50.41 | $-27$:34:00.98 | $-29$:25:23.38 |
| $\ell$ | 3.02 | 2.61 | 0.23 |
| $b$ | $-2.46$ | $-1.72$ | $-1.33$ |
| $(\Gamma_K, \Gamma_O)(hr^{-1})$ | (4, 0) | (4, 1) | (4, 0) |

Here $A_\lambda$ is total magnification, and $f_{i,\lambda}$ is the baseline flux at wavelength $\lambda$ of each source, with $i = 1$ and 2 corresponding to the primary and the secondary sources, respectively.

In addition, we check whether higher-order effects can be constrained from the observed data. The first is the microlensing parallax effect (Gould 1992, 2000, 2004), in which Earth's acceleration around the Sun introduces deviation from rectilinear motion between the lens and the source. We fit it by the inclusion of two parameters $\pi_{E,N}$ and $\pi_{E,E}$, the north and east components of the microlensing parallax vector $\vec{\pi}_E$ in equatorial coordinates,

$$\vec{\pi}_E \equiv \frac{\pi_{rel}}{\theta_E} \hat{\vec{\mu}}_{rel}. \quad (4)$$

We also fit for the $u_0 > 0$ and $u_0 < 0$ solutions for the "ecliptic degeneracy" (Jiang et al. 2004; Poindexter et al. 2005). The second effect is the lens orbital motion (Batista et al. 2011; Skowron et al. 2011), and we fit it by introducing two parameters, $ds/dt$ and $d\alpha/dt$, the instantaneous changes in the separation and orientation of the two lens components defined at $t_0$, for linearized orbital motion. In order to exclude unbound orbits, we restrict the MCMC trials to $\beta < 1.0$ to remove unbound systems, where $\beta$ is the absolute value of the ratio of projected kinetic to potential energy (An et al. 2002; Dong et al. 2009),

$$\beta \equiv \left| \frac{KE_\perp}{PE_\perp} \right| = \frac{\kappa M_\odot yr^2}{8\pi^2} \frac{r^2}{\theta_E} \frac{|\vec{\pi}_E|}{\theta_E} \gamma^2 \left( \frac{s}{|\vec{\pi}_E| + \pi_S/\theta_E} \right)^3; \quad \vec{\gamma} \equiv \left( \frac{ds/dt}{s}, \frac{d\alpha}{dt} \right), \quad (5)$$

and where $\pi_S$ is the source parallax.

## 5.2 KMT-2019-BLG-1042

Figure 2 shows the observed data of KMT-2019-BLG-1042 together with the best-fit models. There is a $\Delta I \sim 0.9$ mag bump 1.2 days before the peak of an otherwise normal PSPL event, and both 2L1S and 1L2S models can produce such an anomaly. Table 2 presents the best-fit parameters and their $1\sigma$ uncertainty for different models. The 2L1S modeling yields two degenerate solutions with $\Delta\chi^2 = 3.7$, and the top panels of Figure 3 display the caustic structures and the source trajectories. The topology of the degeneracy is similar to the

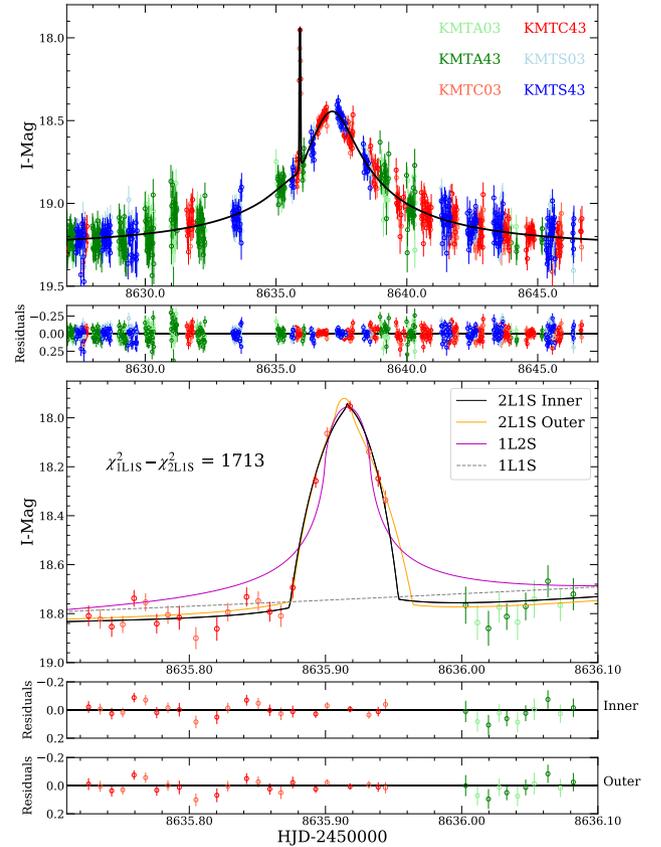

**Figure 2.** Light curve and models for KMT-2019-BLG-1042. The open circles with different colors represent the observed data from different data sets. Different models are shown with different colors. The bottom panels display a close-up of the planetary signal and the residuals to different 2L1S models.

topology of the planetary event, OGLE-2019-BLG-0960 (Yee et al. 2021). The two solutions both have $s > 1$ and a resonant caustic, which is a combination of the central and planetary caustics. The two geometries can be regarded as the source passing "inside" or "outside" the planetary caustics to the central caustic. Thus, we label the two solutions by "Inner" and "Outer", as shown in the top panels of Figure 3. We also find that this event does not suffer from the degeneracy in $\rho$ which occurred in several known planetary events whose planetary signals are also bumps (e.g., Bennett et al. 2014; Ryu et al. 2022). Because the 1L2S model is disfavored by $\Delta\chi^2 = 235$ and cannot fit the bump, we exclude the 1L2S model. In addition, we note

**Table 2.** 2L1S and 1L2S Parameters for KMT-2019-BLG-1042

| Parameters | 2L1S Inner | 2L1S Outer | 1L2S |
|---|---|---|---|
| $\chi^2$/dof | 4162.7/4159 | 4159.0/4159 | 4393.9/4158 |
| $t_{0,1}$ (HJD$'$) | $8637.130 \pm 0.010$ | $8637.135 \pm 0.010$ | $8637.194 \pm 0.012$ |
| $t_{0,2}$ (HJD$'$) | ... | ... | $8635.916 \pm 0.012$ |
| $u_{0,1}$ | $0.083 \pm 0.003$ | $0.076 \pm 0.003$ | $0.067 \pm 0.007$ |
| $u_{0,2}$ | ... | ... | $0.0009 \pm 0.0004$ |
| $q_{f,I}$ | ... | ... | $0.016 \pm 0.001$ |
| $t_E$ (days) | $10.52 \pm 0.33$ | $11.21 \pm 0.36$ | $12.82 \pm 1.21$ |
| $\rho_1$ ($10^{-3}$) | $1.18 \pm 0.17$ | $1.29 \pm 0.21$ | $< 130$ |
| $\rho_2$ ($10^{-3}$) | ... | ... | $1.35 \pm 0.27$ |
| $\alpha$ (rad) | $3.767 \pm 0.010$ | $3.749 \pm 0.010$ | ... |
| $s$ | $1.125 \pm 0.004$ | $1.017 \pm 0.004$ | ... |
| $q$ ($10^{-4}$) | $6.39 \pm 0.64$ | $6.29 \pm 0.60$ | ... |
| $I_S$ | $21.69 \pm 0.04$ | $21.69 \pm 0.04$ | $21.87 \pm 0.10$ |

NOTE. The upper limits on $\rho$ in this paper are all at $3\sigma$ ($\Delta\chi^2 = 9$).

that 2/3 of the KMTA data at HJD$' \sim 8631$(HJD$' =$ HJD$-2450000$) are slightly above the best-fit model by 1–3 $\sigma$. All of these KMTA data were taken under poor seeing ($> 2.6''$), so the weak bump is due to artifacts. However, these data only have a slight impact on the lensing parameters ($< 2\%$).

We find that the inclusion of higher-order effects only improves the fitting by $\Delta\chi^2 < 1$ and the $1\sigma$ uncertainty of parallax at all directions is $> 5$. This is expected based on the relatively short event timescale ($t_E \sim 11$ days) and the short duration of the planetary signal. Thus, we adopt the parameters of the 2L1S model without higher-order effects. This is a new microlensing planet with $q \sim 6.3 \times 10^{-4}$; i.e., about two times the Saturn/Sun mass ratio.

### 5.3 KMT-2019-BLG-1552

As shown in Figure 4, the light curve of KMT-2019-BLG-1552 exhibits two bumps and a trough between them. The 2L1S modeling yields a pair of inner/outer solutions for the minor-image perturbations, for which the source passes on one side of the minor-image planetary caustics or the other (Gaudi & Gould 1997). See Table 3 for their parameters and Figure 3 for their geometries. However, the degeneracy is broken in this case, and the Inner solution is favored by $\Delta\chi^2 = 87$. Gaudi & Gould (1997) argued that the degeneracy would be generally resolvable for the minor-image perturbations, because if the source trajectories approach the two different planetary caustics at different distances in the two solutions, then the bumps of the two solutions would be different. This argument is well-illustrated by the lower panels of Figure 4, in which the main differences for the two solutions come from the two bumps and neither bump can be fitted by the Outer solution[4]. In addition, the Outer solution is still disfavored by $\Delta\chi^2 > 80$ with the inclusion of high-order effects, thus we exclude it. We also exclude the 1L2S model by $\Delta\chi^2 > 1500$.

Due to the long Einstein timescale, $t_E \sim 116$ days, it may be possible to either measure or place a strong constraint on the microlensing parallax vector $\vec{\pi}_E$. As presented in Table 3, the addition of parallax improves the fit by $\Delta\chi^2 \sim 97$. We find that all the seven data sets exhibit $\chi^2$ improvement to the non-parallax model, so the parallax signal is reliable. The component of $\vec{\pi}_E$ that is parallel with the direction of Earths acceleration, $\pi_{E,\parallel}$, is measured by $> 6\sigma$, and the perpendicular component, $\pi_{E,\perp}$, is constraint with $\sigma(\pi_{E,\perp}) < 0.2$. For the lens orbital motion effect, we first fit it without the microlensing parallax and found $\Delta\chi^2 = 27$ worse than the parallax-only model. Then, the inclusion of both microlensing parallax and the lens orbital motion only improves the fit by $\Delta\chi^2 < 1$ compared with the parallax-only models, and $\vec{\gamma}$ has no correlation with $\vec{\pi}_E$. Thus, we conclude that only the microlensing parallax is measured and eliminate the lens orbital motion from the fit. This is a massive microlensing planet with about five times the Jupiter/Sun mass ratio.

### 5.4 KMT-2019-BLG-2974

Figure 5 shows that there is a ~ 4-day dip followed by a ~ 2-day bump around a step at HJD$' \sim 8744$. Such an anomaly is likely due to a minor-image perturbation (e.g., Ranc et al. 2021). That is, the source first passes on the relatively demagnified regions that are flanked by two minor-image planetary caustics and then crosses and/or approaches one of the caustics. The 2L1S modeling yields only one solution, and its parameters and caustic geometry are respectively shown in Table 4 and Figure 3. The best-fit model exhibits two short-duration bumps during the anomaly, which are due to a cusp approach and a caustic crossing with one of the minor-image planetary caustics, respectively. Although the best-fit model shows a caustic crossing feature, due to the lack of data during the crossing, a point-source model is consistent within $1\sigma$ level and the $3\sigma$ upper limit is $\rho < 0.010$.

Although the inclusion of parallax improves the fit by only $\Delta\chi^2 = 1.7$ (see Table 4 for the parameters), the observed data provide a constraint with $\sigma(\pi_{E,\parallel}) \sim 0.2$. We find that the lens orbital motion effect is not detectable ($\Delta\chi^2 < 1$ for 2 degree-of-freedom) and not correlated with $\vec{\pi}_E$. Thus, we adopt the constraint on $\vec{\pi}_E$ in the Bayesian analysis in Section 6.2. The binary mass ratio, $q \sim 6.2 \times 10^{-4}$, indicates that the companion is another microlensing planet with about two times the Saturn/Sun mass ratio.

---

[4] There are, however, many counter-examples, such as OGLE-2012-BLG-0950 (Koshimoto et al. 2017), MOA-2016-BLG-319 (Han et al. 2018), OGLE-2018-BLG-0677 (Herrera-Martín et al. 2020), KMT-2019-BLG-0253, OGLE-2018-BLG-0506, OGLE-2018-BLG-0516, OGLE-2019-BLG-1492 (Hwang et al. 2022), and KMT-2021-BLG-1253 (Ryu et al. 2022).

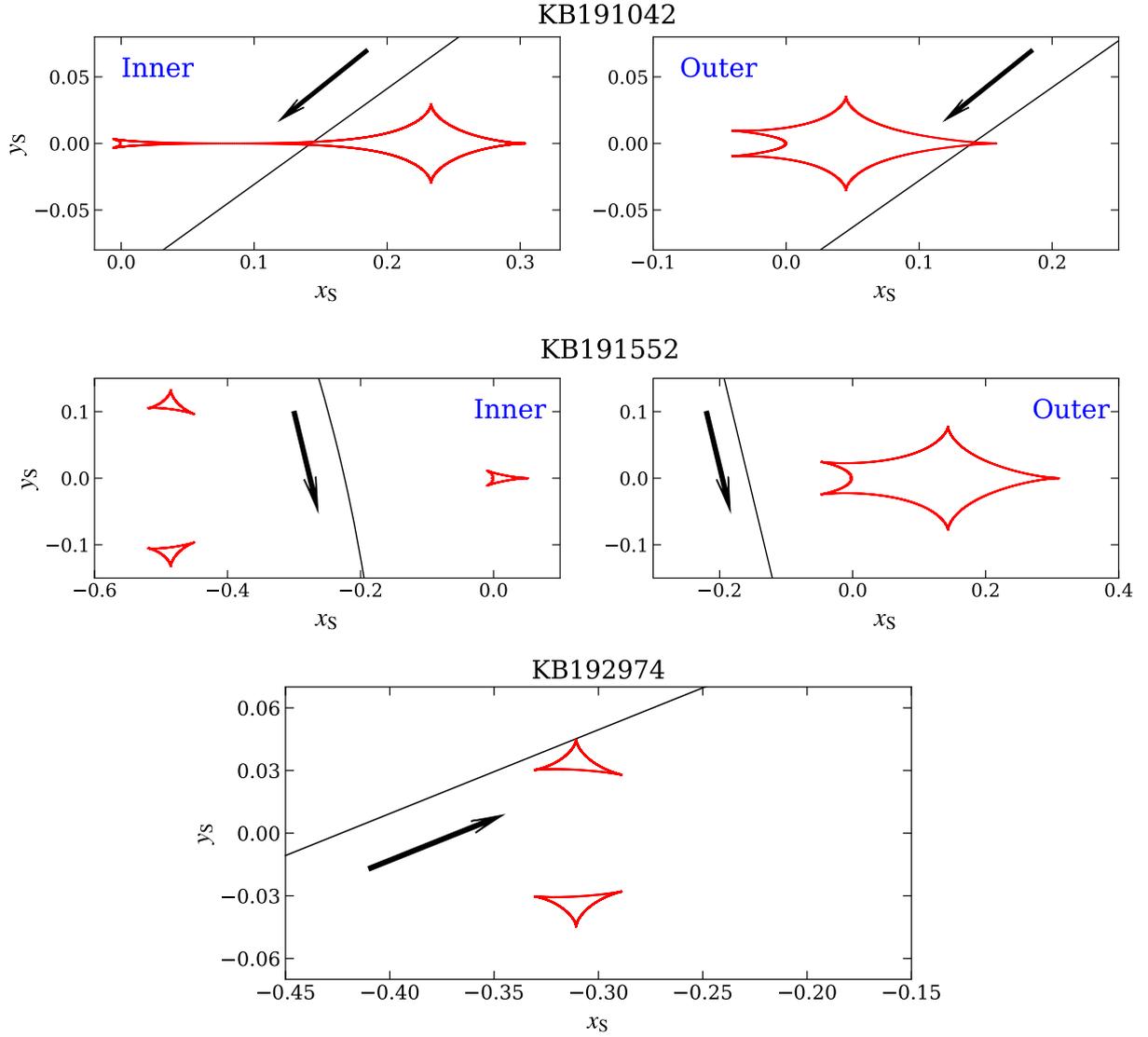

**Figure 3.** Caustic topologies for the three planetary events. Each of the first two events have two degenerate solutions, while the third has a unique solution. In each panel, the red lines represent the caustic, the black line represents the source trajectory, and the line with an arrow indicates the direction of the source-lens relative motion.

**Table 3.** 2L1S Parameters for KMT-2019-BLG-1552

| Parameters | Static (non-parallax) | | Parallax | |
|---|---|---|---|---|
| | Inner | Outer | $u_0 > 0$ | $u_0 < 0$ |
| $\chi^2/$dof | 11736.6/11641 | 11823.7/11641 | 11639.0/11639 | 11640.2/11639 |
| $t_0$ (HJD′) | 8714.249 ± 0.085 | 8714.604 ± 0.076 | 8715.069 ± 0.145 | 8715.019 ± 0.138 |
| $u_0$ | 0.194 ± 0.016 | 0.151 ± 0.016 | 0.215 ± 0.018 | −0.216 ± 0.019 |
| $t_E$ (days) | 116.2 ± 8.2 | 146.5 ± 10.4 | 110.4 ± 9.3 | 106.7 ± 8.3 |
| $\rho$ | < 0.016 | < 0.012 | < 0.017 | < 0.017 |
| $\alpha$ (rad) | 1.819 ± 0.004 | 1.809 ± 0.004 | 1.778 ± 0.013 | −1.789 ± 0.018 |
| $s$ | 0.789 ± 0.005 | 1.074 ± 0.012 | 0.780 ± 0.007 | 0.776 ± 0.006 |
| $q$ $(10^{-3})$ | 4.32 ± 0.22 | 3.74 ± 0.24 | 4.64 ± 0.52 | 5.13 ± 0.59 |
| $\pi_{E,N}$ | ... | ... | 0.110 ± 0.115 | 0.020 ± 0.187 |
| $\pi_{E,E}$ | ... | ... | 0.082 ± 0.013 | 0.092 ± 0.013 |
| $I_S$ | 20.51 ± 0.10 | 20.81 ± 0.13 | 20.36 ± 0.10 | 20.36 ± 0.11 |

**Table 4.** 2L1S Parameters for KMT-2019-BLG-2974

| Parameters | Static | Parallax | |
|---|---|---|---|
| | | $u_0 > 0$ | $u_0 < 0$ |
| $\chi^2$/dof | 4635.1/4635 | 4633.7/4633 | 4633.4/4633 |
| $t_0$ (HJD') | 8753.026 ± 0.056 | 8753.059 ± 0.061 | 8753.059 ± 0.061 |
| $u_0$ | 0.160 ± 0.014 | 0.164 ± 0.017 | −0.165 ± 0.016 |
| $t_E$ (days) | 28.58 ± 1.94 | 27.85 ± 2.64 | 27.70 ± 2.57 |
| $\rho$ | < 0.010 | < 0.011 | < 0.011 |
| $\alpha$ (rad) | 0.386 ± 0.013 | 0.386 ± 0.056 | −0.392 ± 0.059 |
| $s$ | 0.854 ± 0.011 | 0.851 ± 0.013 | 0.851 ± 0.012 |
| $q$ $(10^{-4})$ | 6.18 ± 1.09 | 6.41 ± 1.85 | 6.33 ± 1.96 |
| $\pi_{E,N}$ | ... | −0.091 ± 1.290 | −0.014 ± 1.352 |
| $\pi_{E,E}$ | ... | 0.317 ± 0.249 | 0.336 ± 0.243 |
| $I_S$ | 21.35 ± 0.10 | 21.32 ± 0.12 | 21.31 ± 0.11 |

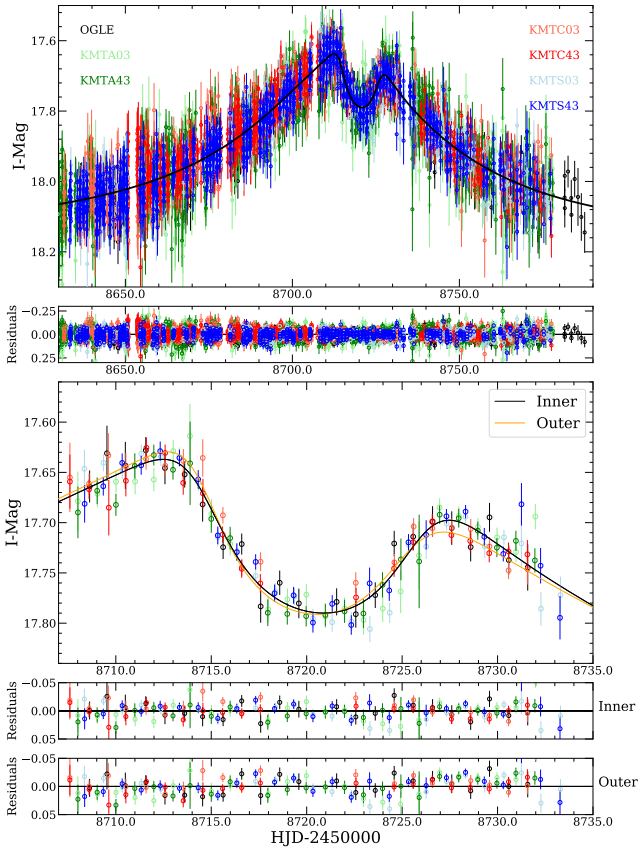

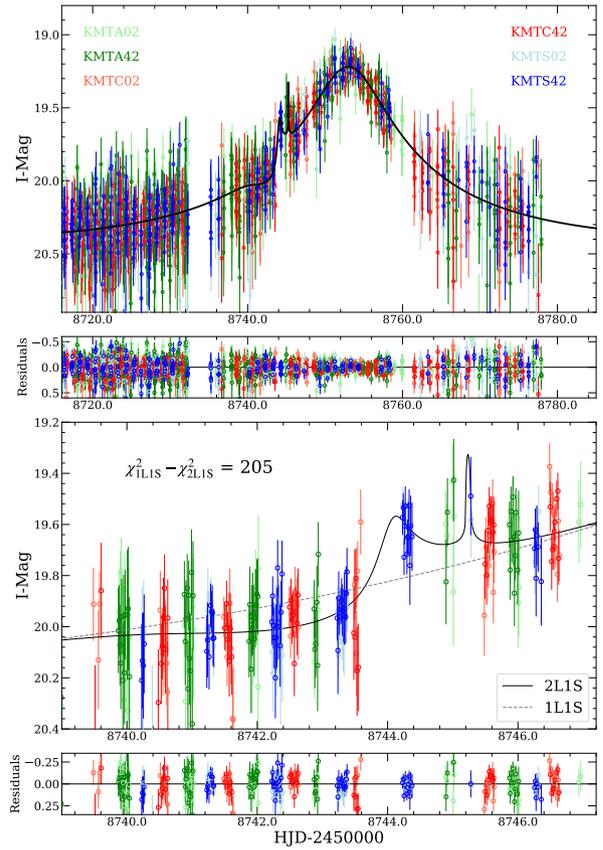

**Figure 4.** The observed data and two 2L1S models for KMT-2019-BLG-1552. The symbols are similar to those in Figure 2. In the bottom panels, for visual clarity, we bin the daily data for each data set.

**Figure 5.** Light curve and models for KMT-2019-BLG-2974. The symbols are similar to those in Figure 2.

## 6  PHYSICAL PARAMETERS

### 6.1  Color Magnitude Diagram

To obtain the angular Einstein radius by $\theta_E = \theta_*/\rho$, we estimate the angular source radius $\theta_*$ by the de-reddened color and magnitude of the source. For each event, we place the source on a color magnitude diagram (CMD, Yoo et al. 2004). For KMT-2019-BLG-1042 and KMT-2019-BLG-1552, the $V - I$ versus $I$ CMDs are constructed using the OGLE-III catalog (Szymański et al. 2011) of stars within a 180″ square centered on the sources. For KMT-2019-BLG-2974, because the $V$-band detection limit of the OGLE-III survey affects the measurement of the red giant clump, the CMD is constructed by a combination of the OGLE-III $I$-band data and the VVV survey (Saito et al. 2012; Minniti et al. 2017) $K$-band data. See Figure 6 for the three CMDs.

We calibrate the pyDIA photometry to the OGLE-III magnitudes. For KMT-2019-BLG-1552, to eliminate the effect from extinction,

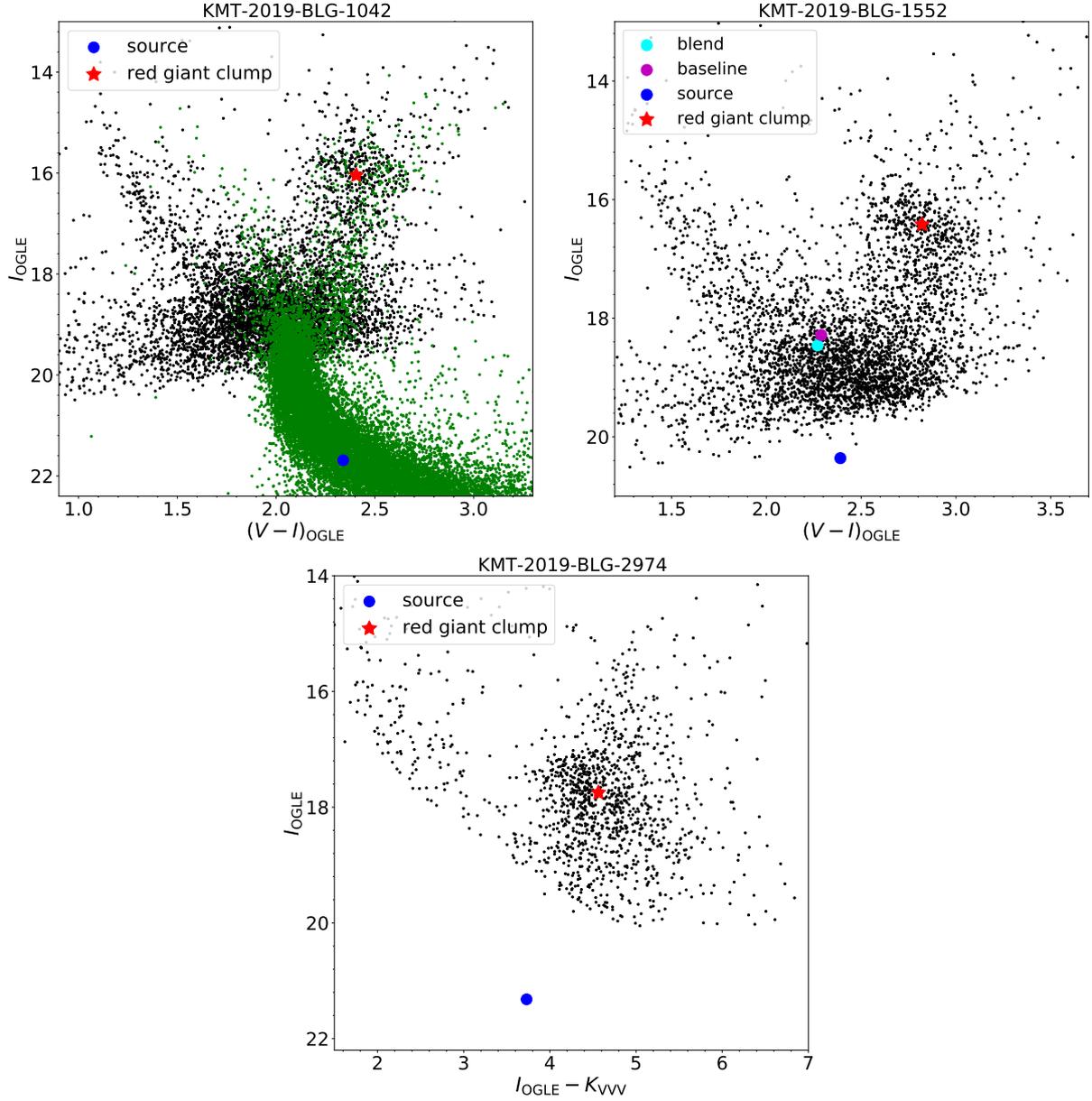

**Figure 6.** Color-magnitude diagrams (CMDs) of a 180″ square centered on each event. For the upper panels, the two CMDs are constructed from the OGLE-III star catalog (Szymański et al. 2011). For the lower panel, the CMD is constructed from the OGLE-III *I*-band data and the VVV survey (Saito et al. 2012; Minniti et al. 2017) *K*-band data. In each panel, the black dots represent the field stars, the red asterisk represents the centroid of the red giant clump, and the blue dot represents the position of the source star. For the upper left panel, we match the HST CMD (green dots, Holtzman et al. 1998) with the OGLE-III CMD using its centroid of red giant clump, $(V-I, I) = (1.62, 15.15)$ (Bennett et al. 2008). For the upper right panel, the magenta and cyan dots show the positions of the baseline object and the blended light, respectively. For the lower panel, the source color, $(I-K)_{\rm S}$, is derived from Bessell & Brett (1988) and the color of the red giant clump.

we compare the source position in the CMD, $(V-I, I)_{\rm S}$, with the position of the centroid of the red giant clump, $(V-I, I)_{\rm cl}$, whose de-reddened properties, $(V-I, I)_{\rm cl,0}$, can be taken from Bensby et al. (2013) and Nataf et al. (2013), respectively. Then, the source de-reddened color and magnitude can be derived by

$$(V-I, I)_{\rm S,0} = (V-I, I)_{\rm S} - (V-I, I)_{\rm cl} + (V-I, I)_{\rm cl,0}. \quad (6)$$

Finally, using the color/surface-brightness relation for dwarfs and subgiants of Adams et al. (2018), we obtain $\theta_*$. For KMT-2019-BLG-1042 and KMT-2019-BLG-2974, due to the low S/N of the *V*-band observations, we estimate $(V-I)_{\rm S,0}$ using the *Hubble* Space

Telescope (*HST*) CMD of Holtzman et al. (1998). We first obtain the *I*-band magnitude of the red giant clump, $I_{\rm cl}$, by the $V-I$ versus $I$, and $I-K$ versus $I$ CMDs, respectively. Then, the de-reddened *I*-band magnitude of the source can be derived by

$$I_{\rm S,0} = I_{\rm S} - I_{\rm cl} + I_{\rm cl,0}. \quad (7)$$

Finally, we estimate $(V-I)_{\rm S,0}$ using *HST* stars whose brightness are within $5\sigma$ of the source star.

We summarize the CMD values and $(\theta_*, \theta_{\rm E}, \mu_{\rm rel})$ in Table 5. Using Table II of Bessell & Brett (1988) and the extinction law of Nishiyama et al. (2009); Nataf et al. (2016), we estimate the $H$- and $K$-bands magnitudes of the source and present the values in Table 5.

## 6.2 Bayesian Analysis

In principle, the lens mass $M_L$ and the lens distance $D_L$ can be determined if the two observables, $\theta_E$ and $\vec{\pi}_E$, are both measured (Gould 1992, 2000),

$$M_L = \frac{\theta_E}{\kappa|\vec{\pi}_E|}; \qquad D_L = \frac{\mathrm{au}}{|\vec{\pi}_E|\theta_E + \pi_S}. \qquad (8)$$

However, neither the three planetary events have simultaneous measurements for the two observables. Thus, we estimate the physical parameters of the planetary systems from a Bayesian analysis using a Galactic model.

We apply the Galactic model and the procedures presented by Zang et al. (2021). In each case, we create a sample of $10^8$ simulated events based on the Galactic model. For each simulated event, we weight it by

$$\omega_{\mathrm{Gal}} = \theta_E \mu_{\mathrm{rel}} \mathcal{L}(t_E) \mathcal{L}(\theta_E), \qquad (9)$$

where $\mathcal{L}(t_E)$ and $\mathcal{L}(\theta_E)$ are the likelihood of its inferred parameters $(t_E, \theta_E)$ respectively derived from the MCMC chain, and the minimum $\chi^2$ for the lower envelope of the ($\chi^2$ vs. $\rho$) diagram and $\theta_*$. For KMT-2019-BLG-1552 and KMT-2019-BLG-2974, we also weight the simulated events by the likelihood distribution of $\vec{\pi}_E$. For each event, we combine the posterior results of the two solutions by their Galactic-model likelihood and $\exp(-\Delta\chi^2/2)$, where $\Delta\chi^2$ is the $\chi^2$ difference from the best-fit solution.

The resulting posterior distributions of the host mass $M_{\mathrm{host}}$, the planet mass $M_{\mathrm{planet}}$, the lens distance $D_L$, the projected planet-host separation $a_\perp$ and the lens-source relative proper motion $\mu_{\mathrm{rel}}$ are listed in Table 6. The Bayesian analysis shows that all the lens systems consist of a cold giant planet beyond the snow line of its host star (assuming a snow line radius $a_{\mathrm{SL}} = 2.7(M/M_\odot)$ au, Kennedy & Kenyon 2008).

## 6.3 Blended light and Future High-resolution Observations

The brightness and the astrometric alignment of the blended light can sometimes provide additional constraints on the lens properties (e.g., Jung et al. 2020a; Yee et al. 2021). Thus, we check the baseline images taken from the 3.6m Canada-France-Hawaii Telescope (CFHT). For the astrometric and baseline brightness measurements, we adopt the CFHT images ($0.45'' <$ seeing FWHM $< 0.55''$) that were taken from 2020 and 2021 in the SDSS-$i'$ band. For the baseline color measurements, we use the CFHT images obtained from the 2016 *K2C9*-CFHT microlensing survey (Zang et al. 2018) with the SDSS-$i'$ ($0.50'' <$ seeing FWHM $< 0.60''$) and SDSS-$g'$ filters ($0.60'' <$ seeing FWHM $< 0.75''$). We use DoPhot (Schechter et al. 1993) to identify fields stars and do photometry and calibrate the photometry to the OGLE-III star catalog.

For KMT-2019-BLG-1042, the KMTNet Alert-Finder system identified an $I = 19.61$ catalog star as the event position. The CFHT images show that the catalog star is about $0.9''$ away from the true position of the source. There is no obvious object at the source position, and the nearest star is an $I \sim 20.0$ star offset from the source position by about $0.33''$. The result from the CFHT images is consistent with the very faint lens ($I \sim 24$) predicted by the Bayesian analysis.

For KMT-2019-BLG-1552, the CFHT images show significant blended light at the source position, with $(V - I, I)_B = (2.27 \pm$

$0.04, 18.46 \pm 0.03)$. We plot the blended light in the CMD (see the upper right panel of Figure 6) and find that the blend belongs to the foreground main-sequence branch. However, the astrometric offset between the source and the blended light is

$$\Delta\theta(N, E) = (73 \pm 5, 99 \pm 5)\mathrm{mas}, \qquad (10)$$

indicating that the majority of the blended light is not the lens light.

For KMT-2019-BLG-2974, the CFHT images cannot identify any star within $1''$ around the source position, indicating that the lens is faint, which is also the prediction of the Bayesian analysis ($I \sim 24.5$).

Future high-resolution observations can potentially unambiguously determine the mass and distance of the three planetary systems. When the source and the lens are resolved by adaptive optics (AO) imaging, the lens light can be measured, which combined with the angular Einstein radius $\theta_E$ and/or the microlensing parallax $\vec{\pi}_E$ can yield the mass and distance (e.g., Batista et al. 2015; Bennett et al. 2015). In addition, such observations can measure the lens-source relative proper motion $\vec{\mu}_{\mathrm{rel}}$ and thus measure the angular Einstein radius by $\theta_E = \mu_{\mathrm{rel}} \times t_E$. We note that Bhattacharya et al. (2018) resolved the source and lens of the event OGLE-2012-BLG-0950 with Keck AO and the *HST* when the lens-source separation is about 34 mas, for which the source and the lens have roughly equal brightness. In Table 6, we present the predicted lens brightness in the $H$ and $K$ bands from the Bayesian analysis using the stellar isochrones of Marigo et al. (2017).

For KMT-2019-BLG-1042, the lens-source relative proper motion is about 10 mas/yr. Considering the Bayesian analysis indicate that the lens is about 1–2 magnitudes fainter than the source in the near-infrared band, the source and the lens could be resolved using Keck AO in 2023. For KMT-2019-BLG-1552 and KMT-2019-BLG-2974, a measurement of the direction of proper motion, $\hat{\mu}_{\mathrm{rel}} = \hat{\pi}_E$, combined with the 1-D measurement from annual parallax, would further constrain $\vec{\pi}_E$ (Ghosh et al. 2004). For KMT-2019-BLG-1552, the source and lens could have a similar brightness, and the predicted lens-source relative proper motion is low, i.e., about 3 mas/yr. It would probably require a 10-year wait for the current ground-based instruments to resolve the source and the lens, at which point AO on 30m class telescopes, with $\sim 3$ times better resolution, is likely to be available. KMT-2019-BLG-2974 probably also requires a 10-year wait for high-resolution observations.

## 7 DISCUSSION

Including the three planetary events analyzed above and the four that are in preparation, there are 14 known planets in the 2019 KMT-Net prime fields. Among them, the planet discovered in the event KMT-2019-BLG-1715 is orbiting a stellar binary (Han et al. 2021a). Kuang et al. (2022) found that the current KMTNet AnomalyFinder and by-eye searches cannot yield a homogeneously selected sample for planets in binary systems, so we exclude this planet from the current 2019 KMTNet prime-field statistical sample. We list the main parameters of the remaining 13 planets in Table 7, ranked ordered by $\log q$. Of these, nine were discovered by the KMTNet AnomalyFinder, and five were first discovered using by-eye searches and then recovered by the KMTNet AnomlyFinder. Among the eight planets with $q < 0.33^4$, only one was first discovered using by-eye searches. That is, the KMTNet AnomalyFinder roughly tripled the planetary detection in the 2019 KMTNet prime fields and played a decisive role in the detection of low-$q$ planets. In addition, except KMT-2019-BLG-1042, the new planets from the KMTNet AnomalyFinder all have

**Table 5.** CMD parameters, $\theta_*$, $\theta_E$ and $\mu_{rel}$ for the three planetary events

| Parameter | KMT-2019-BLG-1042 | | KMT-2019-BLG-1552 | KMT-2019-BLG-2974 |
|---|---|---|---|---|
| | Inner | Outer | | |
| $(V - I, I)_{cl}$ | (N.A., 16.05 ± 0.03) | ← | (2.82 ± 0.01, 16.42 ± 0.02) | (N.A., 17.75 ± 0.02) |
| $(V - I, I)_{cl,0}$ | (1.06 ± 0.03, 14.35 ± 0.04) | ← | (1.06 ± 0.03, 14.36 ± 0.04) | (1.06 ± 0.03, 14.43 ± 0.04) |
| $(V - I, I)_S$ | (N.A., 21.69 ± 0.04) | ← | (2.39 ± 0.05, 20.36 ± 0.10) | (N.A., 21.35 ± 0.10) |
| $(V - I, I)_{S,0}$ | (1.03 ± 0.10, 19.99 ± 0.06) | ← | (0.63 ± 0.05, 18.30 ± 0.11) | (0.74 ± 0.07, 18.03 ± 0.11) |
| $H_S$ | 19.15 ± 0.14 | ← | 18.07 ± 0.13 | 17.95 ± 0.15 |
| $K_S$ | 18.92 ± 0.15 | ← | 17.86 ± 0.14 | 17.62 ± 0.16 |
| $\theta_*$ ($\mu$as) | 0.429 ± 0.045 | 0.429 ± 0.045 | 0.66 ± 0.05 | 0.82 ± 0.06 |
| $\theta_E$ (mas) | 0.364 ± 0.065 | 0.333 ± 0.064 | > 0.039 | > 0.082 |
| $\mu_{rel}$ (mas yr$^{-1}$) | 12.6 ± 2.3 | 10.8 ± 2.1 | > 0.13 | > 1.05 |

NOTE. $H_S$ and $K_S$ are the source magnitudes in the $H$ and $K$ bands.

**Table 6.** Physical parameters of the three planetary events from a Bayesian analysis.

| Name | KMT-2019-BLG-1042 | KMT-2019-BLG-1552 | KMT-2019-BLG-2974 |
|---|---|---|---|
| $M_{host} [M_\odot]$ | $0.30^{+0.34}_{-0.18}$ | $0.79^{+0.33}_{-0.36}$ | $0.47^{+0.41}_{-0.28}$ |
| $M_{planet} [M_J]$ | $0.19^{+0.21}_{-0.11}$ | $4.05^{+1.73}_{-1.85}$ | $0.28^{+0.29}_{-0.18}$ |
| $D_L [kpc]$ | $6.6^{+0.8}_{-1.7}$ | $4.6^{+1.7}_{-1.4}$ | $6.1^{+1.5}_{-2.1}$ |
| $a_\perp [au]$ | $1.7^{+0.5}_{-0.5}$ | $2.6^{+0.6}_{-0.8}$ | $2.0^{+0.8}_{-0.7}$ |
| $\mu_{rel} [mas\,yr^{-1}]$ | $9.1^{+2.2}_{-2.1}$ | $3.2^{+1.8}_{-1.5}$ | $5.7^{+3.6}_{-2.4}$ |
| $H_L [mag]$ | $22.1^{+2.1}_{-2.0}$ | $18.0^{+1.6}_{-1.6}$ | $20.9^{+2.3}_{-2.2}$ |
| $K_L [mag]$ | $21.6^{+2.0}_{-1.9}$ | $17.8^{+1.5}_{-1.5}$ | $20.5^{+2.1}_{-2.1}$ |

NOTE. $H_L$ and $K_L$ are the predicted lens magnitudes in the $H$ and $K$ bands from a Bayesian analysis.

lower $\Delta\chi^2$ than any of the recoveries. This shows that the KMTNet AnomalyFinder is more sensitive than by-eye searches.

Because the 13 planets form a complete statistical sample, we can study their mass-ratio distribution. In Figure 7, we plot the cumulative distributions of $\log q$ for this sample and the 23-planet sample from the MOA-II survey (Suzuki et al. 2016). Hwang et al. (2022) found a roughly uniform $\log q$ distribution over the range $-5.0 < \log q < -3.7$ using the 11 planets from 2018–2019 KMTNet prime fields, and our sample extends the uniform feature to $\log q \sim -1.5$. Compared with the MOA-II sample, our sample has a higher rate of detections (4/13) below the apparent break in the mass-ratio function of the MOA-II sample, $q_{break} = 1.7 \times 10^{-4}$ (Suzuki et al. 2016). Because the KMTNet survey is more powerful than the MOA-II survey in terms of the number of telescopes, cadence, and observational depth, the distributions of the two samples cannot be directly compared and the planetary sensitivity of each sample needs to be considered. However, because the planetary sensitivity function is declining toward lower $q$, our sample suggests that the low-$q$ planets below these $q_{break} = 1.7 \times 10^{-4}$ may be more common than previously believed.

Finally, it should be noted that our current sample is still small and may be affected by small-number statistics. Based on the preliminary search for 2016–2019 KMTNet events, the KMTNet AnomalyFinder is expected to yield a total of 30 to 50 planetary events per year, and a complete analysis of 2016–2019 KMTNet events would yield a sample of about 120 planets, with roughly 20 planets having $q \leqslant 10^{-4}$. This sample, combined with the planetary detection efficiency

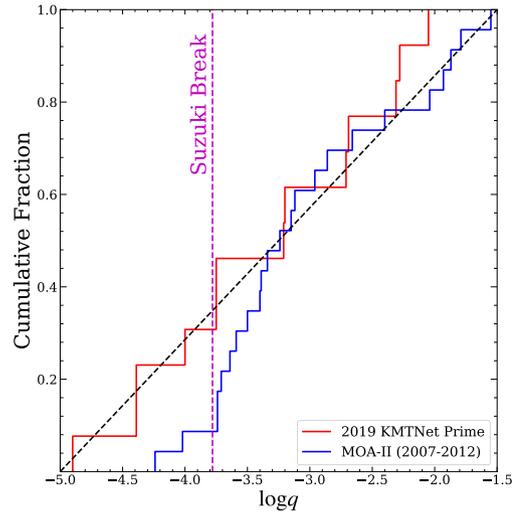

**Figure 7.** Cumulative distribution of $\log q$ for the 2019 KMTNet prime-field planetary sample (red) and the MOA-II planetary sample (blue, Suzuki et al. 2016). The magenta line represents the mass-ratio function break proposed by Suzuki et al. (2016), and the black dashed line simply connects the $\log q = -5.0$ and $\log q = -1.5$. Compared to the MOA-II planetary sample, the 2019 KMTNet prime-field planetary sample is basically uniform in $\log q$ and has a higher rate of detections below the mass-ratio function break.

**Table 7.** Information of the statistical planetary sample from the 2019 KMTNet prime fields

| Event Name | KMTNet Name | $\log q$ | $s$ | $\Delta\chi^2_0$ | Method | Reference |
|---|---|---|---|---|---|---|
| OB191053 | KB191504 | −4.90 | 1.41 | 597.3 | Discovery | Zang et al. (2021) |
| KB190253 | KB190253 | −4.39 | 0.93/1.01 | 440.7 | Discovery | Hwang et al. (2022) |
| KB190842 | KB190842 | −4.39 | 0.98 | 722.5 | Recovery | Jung et al. (2020b) |
| KB191470 | KB191470 | −4.00 | 0.84/1.01 | 529.5 | Discovery | Kuang et al. in prep |
| OB191492 | KB193004 | −3.75 | 0.90/1.04 | 156.7 | Discovery | Hwang et al. (2022) |
| KB190953 | KB190953 | −3.75 | 0.74 | 336.1 | Discovery | Hwang et al. (2022) |
| KB192974 | KB192974 | −3.21 | 0.85 | 297.2 | Discovery | this work |
| KB191042 | KB191042 | −3.20 | 1.02/1.23 | 1710.5 | Discovery | this work |
| KB191953 | KB191953 | −2.71 | 0.40/2.51 | 9488.6 | Recovery | Han et al. (2020) |
| OB190954[a] | KB193289 | −2.69 | 0.71 | 547.3 | Discovery | Kuang et al. in prep |
| KB191552 | KB191552 | −2.31 | 0.78 | 3987.2 | Recovery | this work |
| OB191180 | KB191912 | −2.44 | 1.87 | 19821.1 | Recovery | Chung in prep |
| OB190954[a] | KB193289 | −2.05 | 0.80 | 1092.3 | Recovery | Han et al. (2021b) |

NOTE. Event names are abbreviations, e.g., OGLE-2019-BLG-1053 to OB191053. All $(\log q, s)$ from "in prep" are preliminary values. "Discovery" means that the planet was discovered by the KMTNet AnomlyFinder, and "Recovery" represents that the planet was first discovered via by-eye searches and then recovered by the KMTNet AnomlyFinder. $a$: Two-planet system.

calculator (Jung et al. in prep), will eventually determine whether there is a break in the mass-ratio distribution.

## ACKNOWLEDGEMENTS

We appreciate the anonymous referee for helping to improve the paper. W.Zang, H.Y., S.M. and W.Zhu acknowledge support by the National Science Foundation of China (Grant No. 12133005). This research has made use of the KMTNet system operated by the Korea Astronomy and Space Science Institute (KASI) and the data were obtained at three host sites of CTIO in Chile, SAAO in South Africa, and SSO in Australia. Work by C.H. was supported by the grants of National Research Foundation of Korea (2019R1A2C2085965 and 2020R1A4A2002885). This research is supported by Tsinghua University Initiative Scientific Research Program (Program ID 2019Z07L02017). Work by J.C.Y. acknowledges support from N.S.F Grant No. AST-2108414. W.Zhu acknowledges the science research grants from the China Manned Space Project with No. CMS-CSST-2021-A11. The authors acknowledge the Tsinghua Astrophysics High-Performance Computing platform at Tsinghua University for providing computational and data storage resources that have contributed to the research results reported within this paper. This research has made use of the NASA Exoplanet Archive, which is operated by the California Institute of Technology, under contract with the National Aeronautics and Space Administration under the Exoplanet Exploration Program.

## DATA AVAILABILITY

Data used in the light curve analysis are provided along with publication.

This paper has been typeset from a TEX/LATEX file prepared by the author.